\documentclass[letterpaper,twocolumn,10pt]{article}
\usepackage{usenix,epsfig}

\usepackage{endnotes} 
\usepackage{listings} 
\usepackage{color}
\usepackage{hyperref} 
\usepackage[colorinlistoftodos]{todonotes}
\usepackage{tabularx}
\usepackage{xspace}
\usepackage{enumerate}
\usepackage{graphicx}
\usepackage{booktabs}
\usepackage{xparse}

\NewDocumentCommand{\rot}{O{45} O{1em} m}{\makebox[#2][l]{\rotatebox{#1}{#3}}}%
\begin{document}

\parskip 0pt

\newcommand{\resultperc}{12x} 
\renewcommand{\sectionautorefname}{\S}
\renewcommand{\subsectionautorefname}{\sectionautorefname}
\renewcommand{\subsubsectionautorefname}{\sectionautorefname}

\newcommand{\sysname}{CUP\xspace} 
\newcommand{\falsepos}{0\%}
\newcommand{\falseneg}{0.1\%}

%\newcommand{\mat}[1]{\textcolor{blue}{\textbf{MP:} #1}}
%\newcommand{\dmc}[1]{\textcolor{orange}{\textbf{DM:} #1}}
%\newcommand{\nhb}[1]{\textcolor{red}{\textbf{NB:} #1}}
%\newcommand{\scott}[1]{\textcolor{teal}{\textbf{scott:} #1}}

% uncomment to make your comments disappear
\newcommand{\mat}[1]{}
\newcommand{\dmc}[1]{}
\newcommand{\nhb}[1]{}
\newcommand{\scott}[1]{}

\definecolor{mygreen}{rgb}{0,0.6,0} 
\definecolor{mygray}{rgb}{0.5,0.5,0.5}
\definecolor{mymauve}{rgb}{0.58,0,0.82} 
\lstset{ % 
    language=C,
    backgroundcolor=\color{white},   
    basicstyle=\footnotesize,
    breakatwhitespace=false,
    breaklines=true,
    captionpos=b,                    
    commentstyle=\color{mygreen},    
    deletekeywords={...}, 
    escapeinside={\%*}{*)}, 
    extendedchars=true, 
    frame=single,
    keepspaces=true,
    keywordstyle=\color{blue},
    otherkeywords={*,...},
    numbers=left,
    numbersep=5pt, 
    numberstyle=\tiny\color{mygray}, 
    rulecolor=\color{black},
    showspaces=false,
    showstringspaces=false, 
    showtabs=false, 
    stepnumber=2, 
    stringstyle=\color{mymauve},
    tabsize=2, 
    title=\lstname
}

\date{}

\title{\sysname: Comprehensive User-Space Protection for C/C++}

%\author{
%    {\rm Anonymous submission \# 449}
%}
\author{
    {\rm Nathan Burow}\\
Purdue University
\and
{\rm Derrick McKee}\\
Purdue University
\and
{\rm Scott A. Carr}\\
Purdue University
\and
{\rm Mathias Payer}\\
Purdue University
% copy the following lines to add more authors
% \and
% {\rm Name}\\
%Name Institution
} % end author

\maketitle 

% Use the following at camera-ready time to suppress page numbers.
% Comment it out when you first submit the paper for review.
%\thispagestyle{empty}

\subsection*{Abstract}
Memory corruption vulnerabilities in C/C++ applications enable attackers to
execute code, change data, and leak information. Current memory sanitizers do
not provide comprehensive coverage of a program's data.  In particular, existing
tools focus primarily on heap allocations with limited support for stack
allocations and globals. Additionally, existing tools focus on the main
executable with limited support for system libraries.  Further, they suffer from
both false positives and false negatives.

We present Comprehensive User-Space Protection for C/C++, \sysname{}, an LLVM
sanitizer that provides complete spatial and probabilistic temporal memory
safety for C/C++ programs on 64-bit architectures (with a prototype
implementation for x86\_64).  \sysname{} uses a hybrid metadata scheme
that supports all program data including globals, heap, or stack and maintains
the ABI.  Compared to existing approaches with the NIST Juliet test suite,
\sysname{} reduces false negatives by 10x (\falseneg{}) compared to the
state of the art LLVM sanitizers, and produces no false positives.
\sysname{} instruments all user-space code, including libc and other system
libraries, removing them from the trusted code base.

\section{Introduction} \label{sec:intro}
%%%%%%%%%%%%%%%%%%%%%%%

%\nhb{I'm worried that this is setting people up to expect a bug finding paper in
%    firefox etc which is not what we actually do.  I was just looking for a way
%        to motivate our user space requirement, and the need for a sanitizer}

%What is the problem we are interested in?
Despite extensive research into memory safety techniques~\cite{eternal-war},
exploits of memory corruptions remain common~\cite{memerrors-past, ghost-cve,
getaddrinfo}.  These attacks rely on the fact that C/C++ require the programmer
to manually enforce spatial safety (bounds checks) and temporal safety (lifetime
checks).  As the continuing stream of memory corruption Common
Vulnerabilities and Exposures (CVEs) shows, these programmer added checks are
often inadequate.  Many of these bugs are in network facing code such as
browsers~\cite{firefox-cve, chrome-cve} and servers~\cite{apache-cve,
nginx-cve}, allowing attackers to illicitly gain arbitrary execution on
remote systems. Consequently, a memory safety sanitizer that
\emph{comprehensively} protects user-space is necessary to find and fix
these bugs.

To correctly address memory safety in user-space, there are four main
requirements. \emph{Precision} addresses spatial safety by requiring that exact
bounds are maintained for all allocations.  \emph{Object Awareness}
prevents temporal errors by tracking whether the pointed-to object is currently
allocated or not.  These two requirements are sufficient to enforce memory
safety.  Adding \emph{Comprehensive Coverage} expands this protection to all of
user space by requiring that all data on the stack, heap and globals be
protected.  \emph{Comprehensive Coverage} implies that all code must be
instrumented with the sanitizer, including system libraries like libc.  A
sanitizer that meets these three requirements is powerful enough to find all
memory corruption vulnerabilities in user-space programs.  To be useful, such a
sanitizer must also be usable in practice.  Requiring \emph{Exactness} -- no
false positives and minimal false negatives -- ensures that bugs reported by a
sanitizer are real, and that all spatial and most temporal violations are found.
We discuss these challenges in~\autoref{sec:design}.

%prior work
The research community has come up with many approaches that attempt to address
memory safety.  Initial efforts to address spatial safety relied on ``fat
pointers'' that store bounds information inline with pointers~\cite{ccured,
cyclone}, which unfortunately breaks the Application Binary Interface (ABI).  
SoftBound~\cite{SoftBound.proc} moves metadata to a disjoint table, maintaining 
the ABI, but adding overhead to lookup the metadata associated with tables.
Low-Fat Pointers~\cite{lowfat} reduces overhead by partitioning and
aligning memory allocations, allowing alignment based bounds checks. This
however rounds object sizes, and alters the memory layout of the program.
AddressSanitizer~\cite{addresssanitizer} (ASan) provides probabilistic
spatial safety by relying on poison zones between objects, but
is vulnerable to ``long strides'' that skip these zones.
There are two main approaches to temporal safety.  Probabilistic
approaches~\cite{diehard, dieharder} change the memory allocator to reduce the
frequency with which memory is reallocated.  Deterministic
schemes~\cite{CETS.proc, dangnull} either maintain per pointer metadata that
knows whether an object is still allocated, or invalidate all pointers when the
object is deallocated.

%open problems
\emph{Comprehensive Coverage} is largely an open problem, with prior work mostly
ignoring the stack and neglecting support for system libraries like libc.
Low-Fat Pointers~\cite{lowfat} is the only work to protect the stack -- 
providing only spatial safety.  No existing work provides spatial and temporal
safety comprehensively for all user-space data (stack, heap, globals) and code
(program code, libc, libraries).  Doing so requires a large amount of additional
metadata to protect the extra allocations~\autoref{ss:usage}, which existing
metadata schemes are unable to handle.  Further, the performance of existing
tools does not allow them to scale to handle the additional memory surface of
the stack and libc.

libc is a particularly critical part of user-space to protect.  It is prone to
memory errors, notably the \texttt{mem*} and \texttt{str*} family of functions
(\texttt{memcpy}, \texttt{strcpy}, \ldots). Memory errors in libc are not
limited to these functions, however as shown by, e.g., GHOST~\cite{ghost-cve}
and a stack overflow in \texttt{getaddrinfo}~\cite{getaddrinfo}.  Consequently,
the \emph{entire} libc needs to be protected, not just certain interfaces.

%Hype our evaluation some
\emph{Exactness} shows how well a memory safety solution protects against
vulnerabilities in practice.  The U.S. National Institute of Standards and
Technology (NIST) maintains the Juliet test suite.  Juliet consists of thousands
of examples of bugs, grouped by class from the Common Weakness Enumeration
(CWE). Juliet reveals that existing, open source memory safety
solutions~\cite{addresssanitizer,SoftBound.proc,CETS.proc} have both false
positives and a non-trivial number of false negatives~\autoref{ss:juliet}. 

%introduce our system
\sysname satisfies all four requirements for a powerful, usable memory
sanitizer.  We introduce a new hybrid metadata scheme which is capable of
storing and using \emph{per object} metadata for the stack, libc, heaps, and
globals. Our metadata is precise and does not require altering the program's
memory layout. Additionally, we introduce a new way to check bounds that
leverages hardware to increase our check's performance.  Hybrid metadata
allows us to meet the \emph{Precision} and \emph{Object Awareness} requirements.
\sysname presents a novel use of escape analysis to reduce the amount of
stack allocations without loss of protection. This reduction allows scaling our
mechanism to include all user-space data, satisfying the \emph{Comprehensive
Coverage} requirement.  Further, \sysname successfully handles all system
libraries, including libc, the first memory sanitizer to do so. Our
evaluation on Juliet~\autoref{ss:juliet} shows that we have %\falsepos 
no false
positives and \falseneg false negatives, considerably advancing the state of the
art for \emph{Exactness}.

\nhb{Pull in actual performance numbers}

%\subsection*{Contributions} 
We present the following contributions:
\begin{itemize}
    \item A new hybrid metadata scheme capable of tracking any
        runtime information about object allocations, and show how it can be 
          applied to memory safety.
    \item The first sanitizer to fully protect user-space, including libc
    \item A new static analysis for determining what stack variables
      require active protection, and present a local protection scheme for
      non-escaping stack variables
%    \item We show how software defined bounds checks can be enforced by
%      hardware, a significant performance enhancement
    \item Evaluation of a \sysname prototype that, using our hybrid metadata 
        model, results in (i) %\falsepos{} 
        no false positives and \falseneg{} false
        negatives on the NIST Juliet C/C++ test suite and (ii) reasonably low
        overhead (in line with other sanitizers).
\end{itemize}

\section{Background and Challenges} \label{sec:background}
%%%%%%%%%%%%%%%%%%%%%%%%%%%%%%%%%%%

Our requirements for \emph{Precision} and \emph{Object Awareness} are designed
to enforce spatial and temporal memory safety, which we define here and then use
to introduce the notion of a capability ID.

\paragraph{Spatial Vulnerabilities} also known as bounds-safety violations --
are over- or under-flows of an object.  Over/under-flows occur when a pointer is
incremented/decremented beyond the bounds of the object that it is currently
associated with.  Even if the out-of-bounds pointer still points to a valid
object, it does not have the capability for the referenced object, and the
operation results in a spatial memory safety violation.  However, this violation
is \emph{only} triggered on a dereference of an out-of-bounds pointer.  The C
standard specifically allows out-of-bounds pointers to exist.  

\paragraph{Temporal Vulnerabilities} also known as lifetime-safety violations --
occur when the object that a pointer's capability refers to is no longer
allocated and that pointer is dereferenced.  For stack objects, this is because
the stack frame of the object is no longer valid (the function it was created in
returned); for heap objects, this happens as a result of a free.  These
errors do not \emph{necessarily} cause segmentation faults (accesses to
unmapped memory), because the memory may have
been reallocated to a new object.  Similarly, we cannot simply track what memory
is currently allocated, because the object at a particular address can change,
which still results in a temporal safety violation. Temporal bugs are at the
heart of many recent exploits, e.g., for Google Chrome or Mozilla Firefox as
shown in the pwn2own contests~\cite{pwn2own}

Violating either type of memory safety can be formulated as a capability
violation.  In our terminology, an object is a discrete memory area, created by
an allocation regardless of location (stack, heap, data, bss under the Linux ELF
format).  A capability identifies a specific object, along with information
about its bounds and allocation status. Pointers retain a capability ID that
identifies the capability of the object that was most recently
assigned -- either directly from the allocation or indirectly by aliasing
another pointer~\cite{mike-hicks}.  Capabilities form a contract, upon
dereference: (i) the pointer must be in bounds, and (ii) the referenced object
must still be allocated. Violating the terms of this contract leads to spatial
or temporal memory safety errors respectively.

\subsection{Vulnerable Objects}\label{ss:objects}
%%%%%%%%%%%%%%%%%%%%%%%%%%%%%%%%%%

%Introduce concept of filtering here
\emph{Comprehensive Coverage} requires that we protect all memory objects.
However, some objects are inherently safe, so it is sufficient to protect all
vulnerable objects. In particular, an object is vulnerable if accesses to it are
calculated dynamically, and not by fixed offsets.  This happens with pointers,
and array accesses (which are just syntactic sugar for pointer
arithmetic).  Dynamic address calculation does not happen for variables on the
stack which are not arrays.  Such local variables are accessed by fixed offsets,
calculated at compile time, from the stack frame, and can only be used
maliciously \emph{after} an initial memory corruption.  In particular,
accessing a pointer on the stack does not need to be protected, only the
pointer's dereference, which dynamically looks up memory.  

%\subsection{Examples}
%%%%%%%%%%%%%%%%%%%%%%%%%
%
%\autoref{lst:memerrors} contains classic examples of both spatial and temporal
%memory safety vulnerabilities.  The function \texttt{spatial()} contains a
%buffer overflow.  The declaration \texttt{char bar[5]} allocates an object of 5
%bytes, and the pointer \texttt{bar} is assigned to that object.
%\texttt{bar[6]} attempts to access the 7\textsuperscript{th} byte associated
%with \texttt{bar}, an overflow of two bytes.  The function \texttt{temporal()}
%contains a temporal memory safety violation.  In line 2, pointer \texttt{foo} is
%associated with a new object of size 5 bytes.  On line 4, this object is
%\texttt{free}d, but \texttt{foo} remains associated with it.  Therefore, on
%line 6 when \texttt{foo} is dereferenced a temporal safety violation occurs.  
%
%\begin{lstlisting}[caption={Memory Safety Errors}, label={lst:memerrors}, 
%                   float={t}] 
%void temporal() { 
%  char *foo = (char*)malloc(5);
%  // Calculations involving foo
%  free(foo);
%  // Further calculations
%  foo[0] = 'a'; 
%} 
%void spatial() { 
%  char bar[5]; 
%  bar[6] = 'b'; 
%}
%\end{lstlisting}

\subsection{Comprehensive Coverage Challenges}\label{ss:usage}
%%%%%%%%%%%%%%%%%%%%%%%%%%%%%%%%%%%%%%%%%%%%%%

\begin{table}[tb]
    \centering
    \begin{tabular}{|r|l|}
        \hline
        Memory Type & Allocations \\ \hline
        global & 0.0006\% \\ \hline
        heap & 0.07\% \\ \hline
        stack & 99.9\% \\ \hline
    \end{tabular}
    \caption{Allocation distribution in SPEC CPU2006.}
    \label{tab:memtypes}
\end{table}

Understanding the scope of the challenge presented by \emph{Comprehensive
Coverage} is critical to understanding \sysname's design. To illustrate this
challenge, we show how intensively programs use different logical regions of
memory. While the operating system presents applications with a contiguous
virtual memory address space, that address space is partitioned into three
logical groups for data: global, heap, and stack spaces.  Global memory is
allocated at application load-time, and is available for the entire lifetime of
the application, (i.e. global memory is never explicitly deallocated).  Heap
memory is requested by the application through the \texttt{new} operator or a
call to \texttt{malloc}, and is deallocated via the \texttt{delete} operator or
a \texttt{free} call.  Stack memory space is implicitly allocated with function
calls, and is again implicitly deallocated with a \texttt{return} call. 

Stack allocations account for almost all (99.9\%) of memory allocations in
SPEC CPU2006 (see \autoref{tab:memtypes}).  This measurement \emph{includes}
allocations made in libc. Further, the latest data from van der
Veen, et. al.~\cite{memerrors-past, memerrors-past-data} show that
stack-based vulnerabilities are responsible for an average of over 15\% of
memory related CVEs annually since tracking began in November 2002.  By
comparison, heap-based vulnerabilities account for an average of 25\% of memory
related CVEs over the same time period. Given the stack's exploitability and 
prevalence, which stresses memory safety designs, protecting it is a key design
challenge for memory safety solutions.

%NHB: This paragraph doesn't really have a home, and seems like an unnecessary
%distraction.  I'm not aware of anyone proposing to do check on arithmetic, so
%lets avoid that issue.
%Another practical question is whether to perform spatial checks -- marking the
%pointer valid or invalid -- on pointer manipulations, e.g., arithmetic, or
%on pointer dereference.  We measured the number of pointer manipulations and
%dereferences in the SPECCPU 2006 benchmarks, again \emph{including} libc, and
%found 52\% more manipulations than dereferences. Consequently, for performance
%purposes, dereferences should be checked to enforce spatial safety.

\section{Design} \label{sec:design}
%%%%%%%%%%%%%%%%

\sysname{} provides precise, complete spatial memory safety and stochastic
temporal memory safety by protecting all program data, including libc (and any
other library).  Safety is enforced, for all program data, by dynamically
maintaining information about the size and allocation status of all objects that
are vulnerable to memory safety errors.  This information is recorded through
our novel hybrid metadata scheme~\autoref{ss:hybrid}.  A compiler-based
instrumentation pass is used to add code that records and checks metadata at
runtime~\autoref{ss:compiler}. We provide a detailed argument for why our
instrumentation guarantees memory safety in~\autoref{ss:guarantee}.

%\subsection{Requirements}
%%%%%%%%%%%%%%%%%%%%%%%%%

%\sysname meets the four requirements for a powerful, usable memory sanitizer:
A powerful usable memory sanitizer must comply to the following requirements:

\begin{enumerate}[I]
    \item \emph{Precision}.  The solution must enforce exact object bounds,
        ideally without changing memory layout (i.e., spatial safety).
    \item \emph{Object Aware}.  The solution must remember the allocation state 
        of any object accessed through pointer (i.e., temporal safety).
    \item \emph{Comprehensive Coverage}.  The solution must fully protect
        a program's user-space memory including the stack, heap, and globals,
        requiring instrumentation and analysis of all code, including system
        libraries such as libc (i.e., completeness).
    \item \emph{Exactness}. The solution must have no false positives, and any false
        negatives must be the result of implementation limitations, not design
        limitations (i.e., usefulness).
\end{enumerate}

%Why we need metadata
These requirements drive the design of \sysname.  Fully complying with the
\emph{Precision} and \emph{Object Awareness} requirements relies on creating
metadata for all allocated objects. While it is possible~\cite{baggybounds,
lowfat} to do alignment based spatial checks without metadata, these schemes
loose precision, alter memory layout, and cannot support \emph{Object
Awareness}.  \emph{Object Awareness} for temporal checks requires metadata to
either lookup whether the object is still valid~\cite{CETS.proc} or to find all
pointers associated with an object and mark them invalid upon
deallocations~\cite{dangnull}.  Consequently, \sysname{} is a metadata based
sanitizer.

\sysname provides \emph{Comprehensive Coverage}, and in particular protects
globals, the heap, and stack by instrumenting all code, including libc. Our
hybrid metadata scheme scales to handle the required number of
allocations~\autoref{ss:usage}, and our bounds check leverages the x86\_64
architecture~\autoref{p:hardware} to perform the required volume of checks
quickly enough to be usable.  Additionally, our compiler pass is robust
enough to handle libc~\autoref{ss:musl}, making \sysname the first memory
sanitizer to do so.

%Failing closed 
\emph{Exactness} is achieved, in part, %by designing \sysname to 
\emph{failing closed}, making a missing check equivalent to
a failed check.  We modify the initial pointer returned by object
allocation~\autoref{ss:compiler}, and our modification marks it illegal for
dereference.  This modification propagates through aliasing and all other
operations naturally.  Consequently, we \emph{must} check all uses of the
pointer for the program to execute correctly.  Such an approach results in
optimal precision at the cost of higher engineering burden (as shown in
\autoref{sec:impl} and \autoref{sec:eval}), but dramatically reduces
false negatives.  False positives are prevented by maintaining accurate
metadata, and having it propagate automatically.  

\subsection{Hybrid Metadata Scheme} \label{ss:hybrid}
%%%%%%%%%%%%%%%%%%%%%%%%%%%%%%%%%%%

\begin{lstlisting}[caption={Enriched Pointer}, label={lst:enriched}, 
                    float={t}]
struct pointer_fields {
  int32 id;
  int32 offset;
}

union enriched_ptr {
  struct pointer_fields capability;
  void *ptr;
}
\end{lstlisting}

To provide \emph{Precision}, \emph{Object Awareness}, and \emph{Comprehensive
Coverage}, we introduce a new hybrid metadata scheme that lets us embed a
capability ID in a pointer without changing its bit width.  This capability
ID ties a pointer to the capability metadata for its underlying object.
\emph{Precision} is provided by the metadata containing exact bounds for
every object, and by not rearranging the memory layout. \emph{Object
Awareness} results from having a unique metadata entry for each
capability ID. 

Providing \emph{Comprehensive Coverage} requires
assigning a capability ID to all vulnerable objects in order to associate their
pointers with the object's capability metadata.  However, the capability ID
space is fundamentally limited by the width of pointers.  To address this limit,
we allow capability IDs to be reused.  Consequently, our capability ID
space only needs to support the maximum number of \emph{simultaneously}
allocated objects.  This allows \sysname to \emph{comprehensively cover}
globals, the heap, and stack for all allocations in long running
applications.

Our metadata scheme that draws inspiration from both fat-pointers
and disjoint metadata (and is thus a ``hybrid'' of the two) for 64-bit
architectures.  We conceptually reinterpret the pointer as a structure with two
fields, as illustrated by~\autoref{lst:enriched}. The first field contains the
pointer's capability ID.  The second field stores the offset into the object.
% of the pointer. (i.e., 0 for the first byte, 5 for the 6th byte, etc.).  
This does not change the size of the pointer, thus maintaining the ABI.
Further, when pointers are assigned, the capability ID automatically transfers
to the assigned pointer without further instrumentation.  

Hybrid metadata rewrites pointers to include the capability ID of their
underlying object and current offset, creating \emph{enriched} pointers.
The size of the offset field limits the size of supported object allocations.
The tradeoffs of the field sizes and our
implementation decisions are discussed in~\autoref{ss:meta-impl}.  While we use
it for memory safety, this design allows access to arbitrary metadata, and 
could be applied for, e.g., type safety, or any property that requires runtime
information about object allocations.

Capability IDs in our hybrid-metadata scheme are indexes into a metadata table.
Each entry in this table is a tuple of the \emph{base} and \emph{end} addresses
for the memory object, required for spatial safety checks.  Each
object that is currently allocated has an entry in the table, leading to a
memory overhead of 16 bytes per allocated object. Note that we do not require
per-pointer metadata due to our hybrid scheme. To reduce the number
of required IDs to the number of \emph{concurrently} active objects, we allow
capability IDs to be reused.  Allowing ID reuse thus allows us to protect long
running programs, as our limit is on concurrent pointers, not total allocations
supported.  The security impact of ID reuse is evaluated
in~\autoref{sec:discussion}.
%However, reusing IDs prevents us from fully guaranteeing temporal
%safety. Nonetheless, an undetected temporal violation is highly unlikely.  

%Paragraph on what would be necessary for a temporal violation to be undetected
The metadata table provides strong probabilistic \emph{Object Awareness}.  
For a temporal safety violation to go undetected, two conditions must hold.
First, the capability ID must have been reused. %, because we invalidate
%the associated metadata when an object is deallocated.  
Second, the accessed memory must be within the bounds of the new object. Current
heap grooming techniques~\cite{pzero-good-corruption, heap-feng-shui} already
require a large number of allocations to manipulate heap state.  Adding the
requirement that the same capability ID also be used makes temporal violations
harder. \autoref{sec:discussion} contains other suggestions to further increase
the difficulty.

We are aware of two memory safety concerns for hybrid metadta: (i) arithmetic
overflows from the offset to the capability ID, and (ii) protecting the metadata
table.  The first concern is addressed by operating on the two fields of the
pointer separately. By treating them like separate variables -- while
maintaining them as one entity in memory -- we prevent under/over flows from the
offset field modifying the capability ID. The second concern is not relevant for 
\sysname -- if all memory accesses are checked, then the metadata table cannot
be modified through a memory violation.
% it is a sanitizer for use in a testing environment (not an
% adversarial environment), meaning we do \emph{not} have to protect the
% metadata table.

\subsection{Compiler Modifications}\label{ss:compiler}
%%%%%%%%%%%%%%%%%%%%%%%%%%%%%%%%%%%

%intro
Our compiler adds instrumentation to create entries in the metadata
table and perform runtime checks.  In particular, we first analyze all
allocations (\emph{Comprehensive Coverage}), and filter them to the ones we must
protect to provide \emph{Precision} and \emph{Object Awareness}. This section
also shows how \sysname \emph{fails closed}, and checks all required pointers,
both of which contribute to its \emph{Exactness}.  

%analysis
% \subsubsection{Analysis}\label{ss:analysis}
%%%%%%%%%%%%%%%%%%%%%%%%

The analysis phase identifies when objects are allocated or deallocated,
and when pointers are dereferenced through an
intra-procedural analysis.  All pointers that
are passed inter-procedurally are instrumented using our metadata scheme,
including all heap allocations.  

We manually annotate libc~\autoref{ss:musl} to mark heap allocations.  For
global variables we protect arrays as entries are referenced
indirectly (i.e., with pointer arithmetic).  Similarly for stack allocations,
we only protect arrays, as well as any address taken variable. We leverage
existing LLVM analysis to find address taken variables.  

%Discuss escape analysis here
Our analysis further divides protected stack allocations into (i) escaping and
(ii) non-escaping allocations.  An allocation does not escape if the following
holds: (i) it does not have any aliases, (ii) it is not assigned to the location
referenced by a pointer passed in as a function argument, (iii) it is not
assigned to a global variable, (iv) it is not passed to a sub-function (our
        analysis is intra-procedural excluding inlining), and (v) is not
returned from the function.  For those that escape, we use our usual metadata
scheme so that the bounds information can be looked up in other functions.  For
those that do not escape, we use an alternate instrumentation scheme.

%Discuss our local instrumentation scheme, maybe cite DCI h/t Scott
The optimized instrumentation for non-escaping stack variables scheme creates
local variables with base and bounds information. Since these allocations
are \emph{only} used within the body of the function, we use local
variables for checks instead of looking up the bounds in the metadata table.
This reduces pressure on our capability IDs, helping us to achieve
\emph{Comprehensive Coverage}.  

%nhb: there were too many back to back \emph
All other allocation sites requiring metadata are instrumented to assign the
object the next capability ID and to create metadata (recording its precise
\emph{base} and \emph{end} addresses) -- returning an enriched pointer.
We create metadata at allocation because it is the only time that we are
guaranteed to know the size of the object.

Identifying deallocations for objects is straightforward.  Global
objects are never deallocated over the lifetime of the program.  Heap objects
are explicitly deallocated by, e.g., \texttt{free()} or \texttt{delete}.  Stack
objects are implicitly deallocated when their dominating function returns.
Deallocations are instrumented to mark associated metadata invalid and to
reclaim the capability ID.

Pointer derefences are found by traversing the use-def chain of identified
pointers. Dereferences are analyzed intra-procedurally, so
we include pointers from function arguments (including variadic arguments) and
pointers returned by called functions in the set of allocations for this
analysis.  We instrument dereferences with a bounds check. Note that
the bounds check implicitly checks that the pointer's capability ID identifies
the correct object.  See~\autoref{ss:guarantee} for a discussion of the safety
guarantees.

\sysname also inserts instrumentation to handle int to pointer casts.  These are
commonly inserted by LLVM during optimization, and have matching pointer to int
casts in the same function.  In this case, and any others where we can identify
a matching pointer to int cast, we restore the original capability ID to the
pointer.  If we are unable to find a matching pointer to int cast, we default to
capability ID zero, which is all of user-space.

\subsection{Memory Safety Guarantees}\label{ss:guarantee}
%%%%%%%%%%%%%%%%%%%%%%%%%%%%%%%%%%%%

We discuss how \sysname{} guarantees spatial memory safety and
probabilistically provides temporal safety.  We assume that all code is
instrumented and capability IDs are protected against arithmetic overflow (as
proposed).

For code that we instrument, we keep a capability ID (and thus metadata) for every
memory object that can be accessed via a pointer.  This subset is sufficient to
enforce spatial memory safety.  Objects that are not accessed via pointers are
guaranteed to be safe by the compiler (if you are reading an \texttt{int}, it will always
emit instructions to read the correct 4 bytes from memory). 

Pointers can be used to read or write arbitrary memory. Further, the address
that they reference is often determined dynamically.  Thus, pointers require
dynamic checks at runtime for memory safety guarantees. As defined in
\autoref{sec:background}, memory objects define capabilities for pointers.
These capabilities include the size and validity of the object.  We only create
capabilities when objects are allocated at runtime. Objects can change size due
to, e.g., \texttt{realloc()} calls, in which case we update our metadata
appropriately by changing \emph{base} and \emph{end} to the new
values~\autoref{ss:instr-impl}. Thus, we
always have correct metadata for every object that has been created since the
start of execution.  The metadata for objects that have not been created yet is
invalid by default.

% Pointer propagation
Pointers can receive values in five ways.  First, pointers can be directly
assigned from the memory allocation, e.g., through a call to \texttt{malloc()}.
We have instrumented all allocations to return instrumented pointers.  Second,
they can receive the address of an existing object, via the \texttt{\&}
operator.  We treat this as a special case of object allocation and instrument
it.  Third, pointers can be assigned to the value of another pointer.  As all
existing pointers have been instrumented under the first two scenarios, this
case is covered as well.  Fourth, pointers can be assigned the result of pointer 
arithmetic. This is handled naturally, with our separate loads preventing overflows 
into the capability ID.
The fifth scenario is a cast from an int to a pointer.  This is exceedingly
rare in well written user-space code.  However, the compiler frequently inserts
these operations in optimized code.  As a result, we have to allow these
operations.  We assume that all ints being cast to
pointer were previously a pointers, and thus instrumented.  

So far we have established that all pointers are enriched with capabilities
that accurately reflect the state of the underlying memory object.  Memory safety
violations occur when pointers are dereferenced~\cite{eternal-war}.  We
instrument every dereference to check the pointers capability and ensure that
the dereference is valid. Because each pointer has a capability and each
capability is up-to-date this ensures full memory safety.  

A programs is memory safe before any pointer dereference happens.
We have shown that each type of pointer dereference is protected.
Consequently, every pointer dereference is valid.  Thus, our scheme preserves
memory safety for the entire program.

\section{Implementation} \label{sec:impl}
%%%%%%%%%%%%%%%%%%%%%%%%

We implemented \sysname{} on top of LLVM version 4.0.0-rc1.  Our compiler pass 
is $\approx$2,500 LoC (lines of code), the runtime is another $\approx$300 LoC 
for $\approx$2,800 LoC total. The line count excludes modifications to our libc,
which required only light annotations~\autoref{ss:musl}. Our pass runs after all
optimizations, so that our instrumentation does not prevent compiler
optimizations.  This also reduces the total amount of memory locations that must
be protected, reducing capability ID pressure.

Here we discuss the technical details of how we implemented \sysname in
accordance with our design~\autoref{sec:design}.  We first discuss how our
hybrid metadata scheme is implemented.  Next we present how we find the sets
of allocations and dereferences required by our design.  With the metadata
implementation in mind, we then show how we instrument allocations and
dereferences.  With these details established, we discuss the modifications
required to libc for it to work with \sysname. 

\subsection{Metadata Implementation}\label{ss:meta-impl}
%%%%%%%%%%%%%%%%%%%%%%%%%%%%%%%%%%%%

Our metadata scheme consists of four elements: (i) a table of information, (ii)
a bookkeeping entry for the next entry to use in that table, (iii) a free
list (encoded in the table) that enables us to reuse entries in the table, and
(iv) how to divide the 64 bits in a pointer between the capability ID and offset
in our enriched pointers~\autoref{ss:hybrid}.  Our metadata table is maintained
as a global pointer to a \texttt{mmap}'d region of memory.  Similarly, the next
entry in that table is a global variable known as \emph{next\_entry}.  

By \texttt{mmap}'ing our metadata table, we allow the kernel to lazily allocate
pages, limiting our effective memory overhead.  Further, our ID reuse scheme
reduces fragmentation of our metadata since it will always reuse a capability ID
before allocating a new one. This also helps improve the locality of our
metadata lookups, reducing cache pressure. Alternative reuse schemes with better
temporal security are discussed in~\autoref{sec:discussion}

To implement our capability ID reuse scheme, we update \emph{next\_entry}
using our free list.  The first entry in our metadata table is
reserved~\autoref{p:deref}, so \emph{next\_entry} is initially one.  The free
list is encoded in the \emph{base} fields of each free entry in the table.
These are all initialized to zero.  When an entry is free'd, the \emph{base}
field is set to the \emph{offset} to the next available table entry.  When we
add a metadata entry, \emph{next\_entry} is incremented, and the offset is
added. When an object is deallocated, we have to update the \emph{base} field
for its corresponding capability ID (\emph{ID}) to maintain the free list
correctly.  This requires calculating the \emph{offset} to the next free entry. 
C code illustrating these operations is in~\autoref{lst:free-list}.

\begin{lstlisting}[caption={Free List}, label={lst:free-list}, 
                   float={t}] 
uint_32 next_entry = 1;

//This is done inline, functions are illustrative

void *on_allocation(size_t base, size_t end){
    size_t offset = table[next_entry].base;
    table[next_entry].base = base;
    table[next_entry].end = end;
    uint_32 ret = 0x80000000 & next_entry;
    next_entry = next_entry + offset + 1;
    return (void *)(ret << 32);
}
void on_deallocation(int id){
    table[id].base = next_entry - id - 1;
    table[id].end = 0;
    next_entry = id;
}
\end{lstlisting}

\begin{figure}[t]
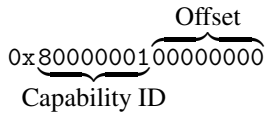
 
    \[
        \texttt{0x}
    \underbrace{\texttt{80000001}}_{\makebox[0pt]{\textrm{Capability ID}}}
    \overbrace{\texttt{00000000}}^{\textrm{Offset}} 
    \]
\caption{A potential pointer value after enrichment.  Note that the high order
    bit is 1 to indicate that it is enriched.} 
    \label{fig:instrument} 
\end{figure}

The final implementation decision for our metadata scheme is how to divide the
64 bits of the pointer between the capability ID and offset.  We use the high
order 32 bits to store our enriched flag and capability
ID~\autoref{fig:instrument}.  This leaves the
low order 32 bits for the offset.  Limiting the offset to 32 bits does limit
individual object size to 4GB under our current design (with up to $2^{31}$ such
allocations).  However, hardware naturally
supports 32-bit manipulations, improving the performance of our implementation.
Further, having a 31-bit capability ID space is crucial for protecting the
entire application~\autoref{ss:usage}.  The enriched bit helps make our
engineering effort easier - with it we can safely check an unenriched
pointer dereferences~\autoref{p:deref}, allowing us to be be conservative and
over-approximate in the set of pointer dereferences that we
check~\autoref{ss:instr-impl}.  The most common use case for this is pointers
returned from the kernel in, e.g., \texttt{malloc()}.  Being able to handle
non-enriched pointers allows us to intervene only when the pointer is returned
to the user from libc~\autoref{ss:musl}, and keep dynamic allocation outside the
trusted computing base.

With a minimal allocation size of 8 bytes, a 31-bit ID allows for at least 16 GB
of allocated memory. In practice, much more memory can be allocated as objects
are usually larger than 8 bytes. When fully allocated, our metadata table uses
\texttt{2GB * sizeof(struct Metadata)}, see~\autoref{lst:instr-example}. Note
that \sysname only allocates pages for ID's that are actually used. 

\subsection{Compiler Pass}
%%%%%%%%%%%%%%%%%%%%%%%%%%

Our LLVM compiler pass operates in two phases: (i) analysis, and (ii)
instrumentation.  As per our design, the analysis phase first determines a
set of code points that require us to add code to perform our runtime checks,
and the instrumentation phases adds these checks.  These checks have been
optimized to let the hardware detect bounds violations rather than doing
comparisons in software. \autoref{lst:instr-example} has a running example for
stack objects.

\begin{lstlisting}[caption={Instrumentation Example}, label={lst:instr-example}, 
                   float={t!}] 
struct Metadata{
    size_t base;
    size_t end;
}

struct Metadata *table;

//The following code is purely illustrative.  
//This is all done inline in LLVM IR in
//our implementation.

static inline size_t check_bounds(size_t base, size_t end, size_t check){
    size_t valid = (check - base) | (end - (check + size));
    valid = valid & 0x8000000000000000;
    return valid;
}

static inline void *check(void *ptr, uint_32 size){
    size_t tmp = (size_t)ptr;
    size_t mask = ptr >> 63;
    uint_32 id = (tmp >> 32) & 0x7fffffff;
    id = id & mask;
    size_t base = table[id].base;
    size_t end = table[id].end;
    size_t check = base + (uint_32)ptr;
    return (void *)(check_bounds(base, end, check) & check);
}

void set(int *x, int val){
    *(check(x, 4)) = val;
}

//example of dereferencing an escaping and a local stack array
int main(void){
    int escapes[5];
    escapes = on_allocation(escapes, escapes+5*4);
    set(escapes[2], 10);

    int local[5];
    size_t local_base = local;
    size_t local_end = local+5*4;
    *(local & check_bounds(local_base, local_end, local+2*4)) = 10;

    on_deallocation(escapes);
}
\end{lstlisting}

\subsubsection{Analysis Implementation}\label{ss:analysis-instr}
%%%%%%%%%%%%%%%%%%%%%%%%%%%%%%%%%%%%%%%

%How we find Allocations
The first task of our pass is to find the set of object allocations that we must
protect to guarantee spatial safety~\autoref{sec:design}.  Heap-based
allocations via \texttt{malloc} are found by our instrumented musl libc as
detailed in \autoref{ss:musl}.  Stack-based allocations are found by examining
\texttt{alloca} instructions in the LLVM Intermediate Representation (IR).  
These are used to allocate all
stack local variables.  However, as detailed in \autoref{ss:objects} we only
target allocations which can be indirectly accessed via, e.g., pointers.
In practice, this means that we need to protect arrays and address-taken
variables on the stack, all others are accessed via fixed offsets from the frame
pointer. Arrays are trivially found by checking the type of \texttt{alloca}
instruction as LLVM's type system for their IR includes arrays.  LLVM's IR has
no notion of the \texttt{\&} operator. However, clang (the C/C++ front end)
inserts markers -- \texttt{llvm.lifetime.start} -- which we use to identify
stack allocations that have their address taken. 

\sysname also protects global variables.  As shown in~\autoref{ss:objects}, we
only need to protect global arrays (logic is identical to protecting stack
arrays).  Global arrays present a challenge for our instrumentation scheme.
\sysname{} relies on changing pointer values. Unfortunately in C/C++ it is
illegal to assign to a global array once it has been allocated.  This means that
we cannot change the pointer's value.  To address this challenge, we create a new
global pointer to the first element in the array, and instrument that
pointer.  We then replace all uses of the global array with our new
pointer that can be manipulated as described above. The new pointer must be
initialized at runtime, once the address of the global array it replace has been
done this.  To do this, we add a new global constructor that initializes our
globals.  The constructor is given priority such that it runs before any code
that relies on our globals.

%Dereference
Once the analysis pass has identified the set of allocations that need to be
protected, it must find all dereferences of pointers to those objects.  For stack
and heap variables, it uses an intra-procedural analysis to do this. In
particular, the analysis pass iterates over the set of protected stack
allocations in the function, pointers passed in as arguments (including variadic
arguments), and pointers returned by called functions.  For each such
pointer, it traces through the use-def chain looking for dereferences. In LLVM
IR, this corresponds to \texttt{load} and \texttt{store} instructions for read
and write derefences respectively. Once these are found, they are added either
to the list of local checks (if the originating allocation is non-escaping), or
the list of checks using our metadata that need to be inserted.

Global variables are handled slightly differently.  LLVM maintains their use-def
chains at the Module level, corresponding to a source file.  Therefore we apply
the same analysis used for stack / heap variables, but it operates over the
entire module instead of intra-procedurally.

\subsubsection{Instrumentation}\label{ss:instr-impl}
%%%%%%%%%%%%%%%%%%%%%%%%%%%%%%%

Our compiler is required to add two types of instrumentation to the code: (i)
allocation, and (ii) dereference.  Allocation instrumentation is responsible for
assigning a capability ID to the resulting pointer, creating metadata for it,
and returning the enriched pointer.  A subcategory of allocation instrumentation
is handling deallocations -- where metadata must be invalidated and the free
list updated.  Dereference instrumentation is responsible for performing our
bounds check, and returning a pointer that can be dereferenced.  While our
runtime library provides functions for the functionality described in this
section (for use in manual annotations), all of our compiler added checks are
done inline.  \autoref{lst:free-list} and \autoref{lst:instr-example} contain
examples for these operations for stack based allocations.

\paragraph{Allocation Instrumentation}
%%%%%%%%%%%%%%%%%%%%%%%%%%%%%%%%%%%%%%

\sysname requires us to rewrite the pointer for every allocation that we
identify as potentially unsafe.  Our rewritten or ``enriched'' pointer contains 
the assigned capability ID, has the enriched bit set, and the lower 32 bits
(which encode the distance from the \emph{base} pointer) are set to 0.  All uses
of the original pointer are then replaced with the new, instrumented pointer. At
allocation time, we use the \emph{next\_entry} global variable as the capability
ID, and then update \emph{next\_entry} as described in~\autoref{ss:meta-impl}.
See the \texttt{escapes} variable in \texttt{main()}
in~\autoref{lst:instr-example}.

Note that this creates a pointer which cannot be dereferenced on x86\_64, which
requires that the high order 16 bits all be 1 (kernel-space) or 0 (user-space).
As a result, any dereference without a check will cause a hardware fault.
Consequently, for any program that runs correctly we can guarantee that all
pointers to protected allocations are checked on dereference.  This is in contrast to
other schemes~\cite{SoftBound.proc} that fail open, i.e., silently continue,  when a
check is missed, sacrificing precision.

When an object is deallocated, we insert code to update the free list in our
metadata table as per~\autoref{ss:meta-impl} and~\autoref{lst:free-list}.  
Further, we mark the \emph{end} address 0 to invalidate the entry.

\paragraph{Dereference Instrumentation}\label{p:deref}
%%%%%%%%%%%%%%%%%%%%%%%%%%%%%%%%%%%%%%%

To dereference a pointer, two things need to be done.  First, we need to
recreate the unenriched pointer.  Then, using the metadata from the enriched
pointer's capability ID, we need to make sure that the unenriched pointer is in
bounds.

To create the unenriched pointer, we first retrieve the high order bit (which
indicates whether the pointer is enriched or not).  We create a 64-bit mask with
the value of this bit.  We then extract the capability ID, and \texttt{AND}
it with this mask.  If the pointer was \emph{not} enriched, this yields an ID of
0 and otherwise preserves the capability ID.  We then lookup the \emph{base}
pointer for the capability ID, and add the offset to it. See \texttt{check}
in~\autoref{lst:instr-example}.

In the case where the pointer was not enriched, we lookup the reserved entry 0
in our metadata table.  This entry has \emph{base} and \emph{end} values that
reflect all of user-space (0 to 0xffffffffffff). Thus, performing our
unenrichment on an unenriched pointer has no effect, and our spatial check below
simply sandboxes it in user-space.

Our spatial check performs the requisite lower and upper bounds check. 
Note, however, that on the upper bound we have to adjust for the
number of bytes being read or written.  This adjustment adds significantly to
our improved precision against other mechanisms~\autoref{ss:juliet}. To
illustrate its importance consider the following.  An \texttt{int} pointer is
being dereferenced, meaning the last byte used is the pointer base + 4 bytes
- 1, while for a \texttt{char} pointer, the last byte used is the pointer base +
1 byte - 1. \autoref{eq:spatial} shows our bounds check formula, the size of 
the pointer dereferenced is \texttt{element\_size}. 

\begin{equation}
    \label{eq:spatial}
    base \leq ptr + element\_size - 1 < base + length
\end{equation}

\paragraph{Hardware Enforcement}\label{p:hardware}
%%%%%%%%%%%%%%%%%%%%%%%%%%%%%%%%

The check in~\autoref{eq:spatial} could naively be implemented with comparison
instructions and a jump -- resulting in additional overhead.  We propose a
novel way to leverage hardware to perform the check for us. We observe that the
difference between the adjusted pointer and the \emph{base} address should
always be greater than or equal to 0.  Similarly, the \emph{end} address minus
the adjusted pointer should always be greater than or equal to 0.  Consequently,
the high order bit in the differences should always be 0 (x86\_64 with two's
complement arithmetic).  We \texttt{OR} these two differences
together, mask off the low order 63 bits, and then \texttt{OR} the result
into the unenriched pointer.  If it passed the check, this changes nothing.
If it failed the check, it results in a invalid pointer, causing a segfault when
dereferenced. \autoref{lst:instr-example} shows this optimized check in
\texttt{check\_bounds()}.

\begin{table*}[t]
   \small
    \setlength{\tabcolsep}{4pt}
    \renewcommand{\arraystretch}{1.2}
    \centering
    %\rotatebox{90} {
        \begin{tabular}{l|rrrrrrrrrrrrrr||r}
              & \rot[90]{\textbf{401.bzip2}} & 
            \rot[90]{\textbf{429.mcf}} & \rot[90]{\textbf{4435.gobmk}}  &
            \rot[90]{\textbf{456.hmmer}}  & \rot[90]{\textbf{458.sjeng}} 
            & \rot[90]{\textbf{462.libquantum}} & \rot[90]{\textbf{464.h264ref}} 
            & \rot[90]{\textbf{473.astar$\dagger$}} & 
            \rot[90]{\textbf{483.xalancbmk$\dagger$}} & \rot[90]{\textbf{433.milc}} 
            & \rot[90]{\textbf{444.namd$\dagger$}}  & \rot[90]{\textbf{450.soplex$\dagger$}}
            & \rot[90]{\textbf{470.lbm}} & \rot[90]{\textbf{482.sphinx3}}
            & \rot[90]{\textbf{Geomean}}
            \\  \hline
            musl-baseline  & 382        & 213   & 386  & 278   &
            371   & 369     & 367     & 327         & 166     &
            431 & 278 & 183      & 201 & 403 & 297 \\
            
            ASan          &       -        &      261     &    -  & 596  &
            630  & 264    & -        &      464       & 369
            & 444  &  377    & 261 & 243 & -  & 369 \\
            
            SoftBound+CETS     & 2004        & -    & -   & -  &
            1372    & -    & -     & -         & -      &
            1543  & -  & -      & 409 & -  & 1148  \\
            
            \sysname{}       &    1131      & 503  &    1255   & 1268 &
            1192  & 391   & 1196   & 955      & -    &
            638  & 905  & -    & 370 & 1185 & 839 \\ 

        \end{tabular}
        %}
    \caption{Performance Results. $\dagger$ indicates C++ benchmark}
\label{tab:perf}
\end{table*}
\subsection{LIBC}\label{ss:musl}
%%%%%%%%%%%%%%%%%

%SCOTT you can not link against libc if you want.  e.g. for microcontrollers
Libc is the foundation of nearly every C program and therefore linked with
nearly every executable.  Unfortunately, many of its most popular functions such
as the \texttt{str*} and \texttt{mem*} functions are highly prone to memory
safety errors.  They make assumptions about program state (e.g., null terminated
strings, buffer sizes) and rely on them without checking that they hold.  Prior
work~\cite{SoftBound.proc, addresssanitizer, lowfat} assumes that libc is part
of the trusted code base (TCB).

In contrast, \sysname{} removes libc from the TCB by instrumenting libc
with our compiler. The majority of the instrumentation is automatic, with few
exceptions such as the memory allocator, system calls, and functions implemented
in assembly code.  The most mature libc implementation that we are aware of that
compiles with Clang-4.0.0-rc1 is musl~\cite{musl}.  Our instrumented musl libc
is part of \sysname{}.

\subsubsection{Dynamic Memory Allocator}
%%%%%%%%%%%%%%%%%%%%%%%%%%%%%%%%%%%%%%%%

The dynamic memory allocator is responsible for requesting memory for the
process from the kernel, and returning it.  To do so efficiently, most
allocators -- including musl -- modify user requests.  In particular, musl
rounds up the number of bytes requested.  Further, musl maintains metadata
inline on the heap in the form of headers before each allocated segment.
Consequently, the allocator's view of memory is different than that of the
program.  

To compensate for this difference, we manually instrumented musl's allocator.
We ensure that pointers are instrumented with the programmer specified length,
not the rounded length.  Further, we manually unenrich pointers as necessary to 
allow musl to read the header blocks that proceed heap data segments.  Without
our intervention, these would appear to be out of bounds.

An interesting corner case is \texttt{realloc()}.  By design it changes the
\emph{end} address.  Additionally, it can change the \emph{base} address if it
was forced to move the allocation to find sufficient room. We manually intervene
in both cases to keep our metadata table up-to-date.

\subsubsection{System Calls}
%%%%%%%%%%%%%%%%%%%%%%%%%%%%

System calls are made through a dedicated API containing
inline assembly in musl.  We initially instrumented this API to ensure that no
enriched pointers are passed to the kernel.  Unfortunately, this is
insufficient as structs containing pointers are passed to the kernel
(\texttt{FILE} structs in particular).  Consequently, we required more context
than the narrow system call API provided.  This forced us to find the actual
system call sites and add additional instrumentation to ensure that all pointers
passed to the kernel are unenriched. 

\subsubsection{Assembly Functions}
%%%%%%%%%%%%%%%%%%%%%%%%%%%%%%%%%%

%functions implemented in asm
musl implements many of the \texttt{mem*}, \texttt{str*} functions in assembly
for x86.  As a result, clang is unaware of these functions as they are directly
assembled and linked in.  We therefore manually instrumented these
functions.  This required minimal intervention to call the relevant function in
our runtime library, while preserving the register state and return value of the
functions we annotated.

\subsubsection{Memory Safety in Musl}
%%%%%%%%%%%%%%%%%%%%%%%%%%%%%%%%%%%%%

%overreads
musl itself is not memory safe.  As an example, \texttt{strlen()} reads 4
bytes at a time as an optimization.  This means that it can overread by as many
as 3 bytes.  To prevent this, we modify \texttt{strlen()} to read byte by byte.
Our experiments show that this does not effect performance for strings of length
less than $\mathcal{O}(10^5)$. Other \texttt{str*} functions overread as well
when looking for the \texttt{NULL} terminator.

\section{Evaluation} \label{sec:eval}
%%%%%%%%%%%%%%%%%%%%

We evaluate \sysname along two axes.  First, we show that its performance is
competitive with existing sanitizers.  Further, our
performance is markedly better than other tools that provide both
\emph{Precision} and \emph{Object Awareness}, given that \sysname provides
\emph{Comprehensive Coverage}. Second, we demonstrate our \emph{Exactness},
showing that we have no false positives and two orders of magnitude fewer
false negatives than existing open source tools on the NIST Juliet suite.
Furthermore, the false negatives that we do register are architecturally
dependent, and do not actually represent a failure to fulfill memory safety
requirements.

All experiments were run on Ubuntu 14.04 with a 3.40GHz Intel Core i7-6700 CPU 
and 16GB of memory. 

\subsection{Performance} \label{ss:perf}
%%%%%%%%%%%%%%%%%%%%%%%%

\begin{table}[tb]
    \centering
    \begin{tabular}{|l|c|}
        \hline
        \sysname{} & 158\% \\ \hline
        ASan & 38\% \\ \hline
        SoftBound+CETS & 53\% \\ \hline
    \end{tabular}
    \caption{Percent Overhead over Baseline}
    \label{tbl:base-overhead}
\end{table}

\begin{table}[tb]
    \centering
    \begin{tabular}{|l|c|}
        \hline
        \sysname{} & 83\% \\ \hline
        SoftBound+CETS & 134\% \\ \hline
    \end{tabular}
    \caption{Percent Overhead over ASan}
    \label{tbl:asan-overhead}
\end{table}

Performance is an important requirement for any usable sanitizer.  We
evaluate the performance of \sysname with musl.  For comparison, we also measure
the performance overhead of similar sanitizers, using glibc as the baseline.
AddressSanitizer is the closest open source related work that is compatible with
LLVM-4.0.0-rc1.  SoftBound+CETS is open source, but relies on LLVM-3.4 and can 
only run a small subset of SPEC CPU2006 benchmarks.  Its performance results are
reported here, but are not directly comparable.  Low-Fat Pointers is not yet
open source.

\autoref{tbl:base-overhead} summarizes the performance results for \sysname.  We
have 158\% overhead vs baseline, compared to 38\% for AddressSanitizer.
Comparing \sysname directly with AddressSanitizer~\autoref{tbl:asan-overhead}
shows that we have 83\% overhead relative to AddressSanitizer. On benchmarks
where both run, \sysname is faster than SoftBound+CETS.  Further, compared to
these existing tools, we offer stronger, more precise
coverage~\autoref{ss:juliet}.  As we show in~\autoref{ss:juliet}, \sysname has
10x the coverage of SoftBound+CETS, for less overhead. Low-Fat
Pointers~\cite{lowfat} reports 16\% to 62\% overhead depending on their
optimization level.  They achieve this by using clever alignment tricks to avoid
metadata look ups.  This has two drawbacks: (i) they round allocations up to the
nearest power of two, losing precision for bounds checks, and (ii) their design
cannot support temporal checks which require metadata.

\subsection{Juliet Suite}\label{ss:juliet}
%%%%%%%%%%%%%%%%%%%%%%%%%

NIST maintains the Juliet test suite, a large collection of C source code that 
demonstrates common programming practices that lead to memory vulnerabilities,
organized by Common Weakness Enumeration (CWE)
numbers~\cite{juliet}.  Every example comes with two versions: one that
exhibits the bug and one that is patched.  We extracted the subset of tests
for heap and stack buffers\footnote{The tests 
that match the following regular expression:
CWE(121$\vert$122)+((?!socket).)*(\textbackslash.c)\$}.  We compiled all sources
with our pass, as well as with SoftBound+CETS\footnote{git commit
\href{https://github.com/santoshn/softboundcets-34/}
{9a9c09f04e16f2d1ef3a906fd138a7b89df44996}} and with
AddressSanitizer~\cite{addresssanitizer}.  Every
patched test should execute normally.  If a memory protection mechanism prematurely
kills a patched test, we call it a false positive.  Conversely, every
buggy test should be stopped by the memory safety mechanism.  All three
memory protection mechanisms kill the process in case of a memory violation. If
the process is not killed, we report a false negative.

\begin{table}[tb]
    \begin{tabular}{|r|c|c|}
        \hline & False Neg. & False Pos. \\ \hline
        \sysname{} & 4 (0.1\%) & 0 (0\%) \\ \hline
        SoftBound+CETS & 1032 (25\%) & 12 (0.3\%) \\ \hline
        AddressSanitizer & 315 (8\%) & 0 (0\%) \\ \hline \hline
        Total Tests & \multicolumn{2}{c|} {4038} \\ \hline
    \end{tabular}
    \caption{Juliet Suite Results}
    \label{tbl:juliet}
\end{table}

\autoref{tbl:juliet} summarizes the results.  Out of 4,038 tests that should not
fail, we incur no false positives.  ASan and SoftBound+CETS 
show a 0\% and 0.3\% respectively.  We produce a 0.1\% false negative
rate, while ASan produces an 8\% false negative rate, and SoftBound+CETS
has a 25\% false negative rate.

The false positives that SoftBound+CETS registers comes from variations in how
the \texttt{alloca()} function call is handled. \texttt{alloca()} is compiler
dependent~\cite{alloca-man-page}.  The failing examples use \texttt{alloca}
which is wrapped around a static function.  SoftBound+CETS uses clang 3.4 as the
underlying compiler, and \sysname{} uses clang 4.0.  Consequently,
SoftBound+CETS handles the examples differently, and sees the memory from
\texttt{alloca} as invalid, while \sysname{} does not.

Our false negatives are architecturally dependent. The examples we fail to catch
attempt to allocate memory for a structure containing exactly two
\texttt{int}s.
However, erroneously, the examples use the size of a pointer to the two 
\texttt{int} structure when allocating memory.
(e.g. \texttt{malloc(sizeof(TwoIntStruct*))}).  On architectures which do not
define pointers as twice the size of \texttt{int}s (including 32-bit x86), 
such a mistake would lead to a memory violation. 
With the x86\_64 architecture, though, the size of a pointer and the size of the 
two \texttt{int} structure are the same.  Thus, while semantically
incorrect, no true memory violation occurs.  No false positives, combined with
no true false negatives shows that \sysname{} fulfills the \textit{Exactness}
requirement.

ASan and SoftBound+CETS higher false negative rate results from
their incomplete \emph{coverage}. In particular, many of the Juliet examples involve 
calling libc functions to copy 
strings and buffers (e.g. \texttt{strcpy} and \texttt{memcpy}).  Neither 
ASan nor SoftBound+CETS are able to protect against unsafe 
libc calls. Our instrumentation of libc~\autoref{ss:musl} allows us to properly
detect memory violations in these calls. Further, our adjustment for read /
write size~\autoref{p:deref} allows us to catch additional memory safety
violations.

\autoref{lst:juliet-example} provides source based on a Juliet
example~\footnote{CWE121\_Stack\_Based\_Buffer\_Overflow\_\_CWE129\_rand\_22\_bad()} 
that \sysname{} properly handles, and which ASan handles incorrectly.  
The code allocates 10 bytes on the stack, 
sets the bytes to ASCII \texttt{A}, and then sets a random character to ASCII
\texttt{B}. ASan protects the stack by surrounding 
stack objects with poison zones.  However, depending on the value of
\texttt{data}, line 5 could be outside the allocated objects \textit{and}
outside the poison region -- a classic buffer overwrite.
\sysname{}'s analysis detects that \texttt{buffer} leaves \texttt{main} through
the call to \texttt{memset}, and inserts the appropriate runtime checks.  If
\texttt{data} is greater than 10, those runtime checks fail.

\begin{lstlisting}[caption={Example of code ASan fails to protect}, 
label={lst:juliet-example}, float={tb}]
int main() {
    int data = rand();
    char *buffer = (char*)alloc(10, 1);
    memset((void*)buffer, 'A', 10);
    buffer[data] = 'B';
}
\end{lstlisting}

\section{Discussion}\label{sec:discussion}
%%%%%%%%%%%%%%%%%%%

We discuss several design aspects of \sysname{}, how we handle specific corner
cases, potential for optimization, and further extensions.

\paragraph{Reducing instrumentation.} Prior work on reducing the amount of
required runtime checks has focused on type systems.  CCured~\cite{ccured}
infers three types of pointers: safe, sequential, and wild.  Safe pointers are
statically proven to stay in bounds.  Sequential pointers are only incremented
(or decremented) -- e.g., iterating over an array in a loop.  All remaining
pointers are classified as wild.  Leveraging CCured-style type systems to
further optimize memory safety solutions is left as an open problem.

%benefits of LTO
\paragraph{Optimization through LTO.} \sysname{} does not depend on Link Time
Optimization (LTO).  However, LTO makes it possible to inline functions across
source files, and generally flattens code.  Inlining would increase the
effectiveness of our stack optimization and further reduce the amount of
instrumented stack variables, reducing the number of IDs that a program
consumes.  Reducing the IDs a program uses, reduces the overhead for ID
management and resources used by \sysname{}. 

\paragraph{Arithmetic overflow.} Our hybrid metadata scheme stores the
capability ID as part of the pointer. Pointer arithmetic can potentially modify
the ID, allowing an adversary to chose the metadata the pointer is checked
against.  To prevent this attack vector, the upper and lower 32 bits of the
pointer are loaded separately.  The compiler enforces that any arithmetic
operations only happen on the lower 32 bits, which contain the pointer's offset.
This protects our capability ID from manipulation by an adversary.  

\paragraph{Stronger temporal protection.} As discussed in \autoref{ss:hybrid},
it is possible (albeit difficult) for an attacker to perform a temporal attack
on software instrumented with \sysname{}.  If \sysname were deployed as an
active defense, the difficulty of a successful temporal attack could be
increased by utilizing a randomized memory allocator like DieHard~\cite{diehard}
or DieHarder~\cite{dieharder}.  Such allocators randomize heap allocations,
making heap grooming~\cite{pzero-good-corruption, heap-feng-shui} much more
difficult.  Beyond getting the addresses to line up, the increased number of
required allocations makes matching the capability IDs even harder. This renders
a successful use-after-free attack highly unlikely, with minimal additional
overhead. Additionally, a ``lock and key'' scheme~\cite{CETS.proc} can be added
to our metadata.  Alternatively, our metadata can be extended to include a
dangnull~\cite{dangnull}-style approach that records how many references are
still pointing to an object and either explicitly invalidating those references
or waiting until the last reference has been overwritten before reusing IDs.

\paragraph{Uninstrumented code.} \sysname{} supports linking with uninstrumented
libraries. Enriched pointers are the same size as regular pointers, maintaining
the ABI. Dereferencing enriched pointers in uninstrumented code results, by
design, in a segmentation fault. A segmentation fault handler can, on demand,
dereference individual pointers. As the memory allocator is instrumented, memory
allocations in uninstrumented code will return enriched pointers. Stack
allocations on the other hand will not be protected in uninstrumented code.
While this option allows compatibility, it clearly results in high performance
overhead. Note that, thanks to support for recompiling all user-space code, this
situation is limited to legacy code.

\paragraph{Uninstrumented globals.} We currently cannot handle the corner case
of external global arrays defined in \emph{uninstrumented} code. Our pass
assumes such arrays have been instrumented, and so adds an extern pointer to
them~\autoref{ss:analysis-instr}.  If the global was defined in uninstrumented
code, this assumption does not hold.  Consequently, our extern pointer does not
exist, and the code will not link. Note that such a situation is unlikely, as
we support full user-space instrumentation.

\paragraph{Assembly code.} \sysname{} automatically instruments all code written
in high level languages; our analysis pass runs on LLVM IR.  Our analysis does
not (and cannot) instrument inline assembly and assembly files due to missing
type information.  We rely on the programmer to either instrument the assembly
code accordingly or to fall back on supporting uninstrumented code as mentioned
above.

\paragraph{Data races.} \sysname{} does not protect against inter-thread races
of updates to metadata (e.g., one thread frees an object while the other thread
is dereferencing a pointer). We leave the design of a low-overhead metadata locking
scheme as future work. Note that this limitation is shared with other
sanitizers.

% \subsection{Uninstrumented Code}
% %%%%%%%%%%%%%%%%%%%%%%%%%%%%%%%%
% 
% Our design of \sysname{} is compatible with uninstrumented code. This allows us
% to support linking against third-party libraries that are not compiled with
% \sysname{}. Our enriched pointers are the same size as normal 64-bit pointers,
% maintaining the ABI.  There are two further considerations: what happens
% when uninstrumented code attempts to dereference an instrumented pointer,
% and what happens when uninstrumented code allocates or deallocates a memory
% object.
% 
% Dereferences of enriched pointers cause segfaults because we have modified the
% upper region of the pointer, and at least the 16 upper bits must all be zero for
% user-space pointers.  Our runtime library handles this for instrumented code
% \autoref{ss:runtime}. For uninstrumented code, a segmentation fault handler can
% catch the illegal access, execute a load or store through the unenriched
% pointer, and resume execution.  While incurring overhead for uninstrumented
% code, this scheme allows us to keep the metadata in the pointer.
% 
% Another issue is uninstrumented code that allocates or deallocates memory. While
% dynamic allocation on the heap is covered through our existing instrumentation
% in libc, stack allocations are challenging as they do not call memory
% management functions but allocate memory by changing the stack pointer (which we
% cannot track easily). Therefore, we cannot guarantee memory safety for
% stack objects in uninstrumented functions.

%%%%%%%%%%%%%%%%%%%%%%
\section{Related Work} \label{sec:related}
%%%%%%%%%%%%%%%%%%%%%%

\paragraph{Precision} is required to enforce spatial memory safety (bounds checks).
There is a class of memory safety solutions that only approximately enforce this
property~\cite{baggybounds, addresssanitizer, lowfat}.  These solutions make use of
techniques such as poisoned zones --  detecting spatial violations within
limits, or rounding allocation size -- which causes the executed program to
differ from the programmers intent and results in challenges when trying to
handle intra-array and intra-struct checks.  By changing the memory layout and not
enforcing \emph{exact} bounds, these solutions are not faithful to the
programmer's intent. \cite{SoftBound.proc} is the existing solution which best
satisfies this requirement, though it has other limitations.

Object-based memory checking~\cite{criswell2007secure,
Dhurjati:2006:BAB:1134285.1134309, eigler2003mudflap} keeps track of metadata on
a per-object basis.  Since the meta-data is associated on a per object level,
every pointer to the object shares the same metadata.  Object layout in memory
is generally left unchanged, which increases ABI compatibility.  However,
pointer casts and pointers to subfield struct members are
unhandled~\cite{pointer_checking}.  SAFECode~\cite{safecode} is an example
efficient object-based memory checking. 

Recently, Intel has started to add memory safety extensions, called Intel MPX,
to their processors, starting with the Skylake architecture~\cite{intelmpx}.
These extensions add additional registers to store bounds information at
runtime.  While effective at detecting spatial memory violations, MPX is
incapable of detecting temporal violations~\cite{mpxeval}, and typecasts to
integers are not protected.  In addition, current implementations of MPX incur a
large memory penalty of up to 4 times normal usage~\cite{mpxexp}.  Other ISA
extensions include Watchdog~\cite{watchdog}, WatchdogLite~\cite{watchdoglite},
HardBound~\cite{hardbound}, and Chuang et al.~\cite{accelerating}.  As a
software only solution, \sysname{} does not require extra hardware or ISA
extensions.

\paragraph{Object Awareness} is required to prevent temporal memory safety violations
(lifetime errors). This property requires remembering for every pointer whether
the object to which it is assigned is still allocated.
AddressSanitizer~\cite{addresssanitizer} and Low-Fat Pointers~\cite{lowfat} make
no attempt to do this (and their metadata does not support this property), while
SoftBound+CETS~\cite{CETS.proc} enforces this property.  AddressSanitizer and
Low-Fat Pointers do not maintain metadata either per object or per pointer.
\emph{Object Awareness} requires either per objecct or per pointer metadata.
Consequently, they fundamentally \emph{cannot} enforce temporal safety because
their (current) metadata \emph{cannot} be object aware.

Temporal only detectors include DangNull~\cite{dangnull} and
Undangle~\cite{undangle} from Microsoft.  DangNull automatically nullifies all
pointers to an object when it is freed.  Undangle uses an early
detection technique to identify unsafe pointers when they are created, instead
of being used.  \sysname{} only provides a probabilistic temporal defense,
however, DangNull and Undangle lack any spatial protection.

\paragraph{Comprehensive Coverage} is required to fully protect the program. As shown
in~\autoref{ss:usage} stack objects are the overwhelming majority of
allocations, and to this day a significant portion of memory safety Common
Vulnerabilities and Exposures (CVE) are stack related. Our evaluation of
SoftBound+CETS~\autoref{ss:juliet} shows that it has poor coverage -- missing
many stack vulnerabilities.  AddressSanitizer and Low-Fat Pointers do better.
AddressSanitizer protects the stack through the use of poisoned zones, and, as
illustrated in ~\autoref{ss:juliet} cannot handle all invalid stack memory
accesses.  Additionally, neither of them supports compiling libc -- leaving the 
window open for vulnerabilities such as GHOST~\cite{ghost-cve}. 
Tripwires~\cite{valgrind, 1385952, 4147668,
Yong:2003:PCP:940071.940113} are a way to detect some spatial and temporal
memory errors~\cite{addresssanitizer, memtracker}. Tripwires place a region
of invalid memory around objects to avoid small stride overflows and
underflows.  Temporal violations are caught by registering memory
freed as invalid, until reclaimed.  Tripwires, however, miss long
stride memory errors, and thus cannot be said to be completely secure.

The state-of-the-art C/C++ pointer-based memory safety scheme is
SoftBound+CETS~\cite{pointer_checking}.  Other pointer-based
schemes include CCured~\cite{ccured} and Cyclone~\cite{cyclone}.  CCured
uses a fat pointer to store metadata, as well as programmer annotations for
indicating safe casts.  Unfortunately, fat pointers break the ABI, and
programmer annotations can significantly increase developer time.  Even with
annotations, CCured fails to handle structure changes.  Cyclone also uses a
fat pointer scheme, but does not guarantee full memory safety. We refer
to~\cite{pointer_checking} for a survey of other related work on memory safety
and a discussion on different pointer metadata schemes.

%\paragraph{Control-Flow Hijack Defenses.} Mechanisms like control-flow integrity
%(CFI)~\cite{cfi, fecvc, picfi} check if a code pointer has been modified to
%point to an illegal address before it is used, e.g., for an indirect function
%call. CFI mechanisms assume that memory safety violations can happen and only
%checks the integrity of code pointers.  Unfortunately, recent research has shown
%that even with fully precise CFI, the existence of libc calls like
%\texttt{printf} allow attacks that do not violate the control flow.  CFI is a
%low-overhead approach to protect programs against illicit uses of a corrupted
%pointer (without protecting the integrity of the code pointer itself) while
%memory safety protects against the pointer corruption in the first place.  CFI
%attacks are fundamentally caused by a memory violation~\cite{eternal-war}, and
%\sysname{}'s \textit{Complete Coverage} fulfillment mitigates such attacks.

\section{Conclusion} \label{sec:concl}
%%%%%%%%%%%%%%%%%%%%%

We present \sysname{}, a C/C++ memory safety mechanism that provides full
user-space protection, including libc, and strong probabilistic temporal
protection.  It is the first such mechanism that satisfies all requirements for
a complete memory safety solution, while incurring only modest performance
overhead compared with the state-of-the-art. \sysname{} is exact, object aware,
comprehensive in its coverage, and precise.  We fully protect all user-space 
memory, including the stack, which, despite being the largest source of pointers,
remained largely unprotected.  Finally, we produce zero false negatives and
zero authentic false positives in the NIST Juliet Vulnerability example suite, 
which represents a significant advancement over existing memory safety
mechanisms.

%We will open source our code upon acceptance.

{\footnotesize \bibliographystyle{acm}
\bibliography{hexsafe}}

\appendix

%\begin{table*}[t]
%    \centering
%   \small
%    \renewcommand{\arraystretch}{1.2}
%        \begin{tabular}{l|llll||l}  
%            \textbf{C++} & \rot[60]{\textbf{473.astar}} &
%            \rot[60]{\textbf{444.namd}} &
%            \rot[60]{\textbf{450.soplex}} & \rot[60]{\textbf{453.povray}}
%            & \rot[60]{\textbf{Geomean}}
%            \\ \hline
%            musl      & 1.26\%    & -50.55\% & 2.34\%     & 0.91\%     &
%            -15.20\% \\
%            asan      & 50.94\%   & 116.27\% & -30.51\%   & 54.97\%    & 36.93\%
%            \\
%            softbound & 4.72\%    & 16.36\%  & 0.58\%     & -0.91\%    & 3.99\%
%            \\
%            hexsafe   & 209.63\%  & 182.90\% &            &            &
%            195.96\%
%        \end{tabular}
%\end{table*}

\end{document}